\documentclass[aps,twocolumn,floats,nofootinbib,pre]{revtex4}
\usepackage{graphics,graphicx,epsfig}
\usepackage{amssymb,color}
\usepackage{epsf,epstopdf,wrapfig}
\usepackage {amsmath}

\newcommand{\beq}{\begin{equation}}
\newcommand{\eeq}{\end{equation}}
\newcommand{\beqn}{\begin{eqnarray}}
\newcommand{\eeqn}{\end{eqnarray}}

\begin{document}

\title{What do we mean by the dimensionality of behavior?}\thanks{This is based in part on a presentation at the Physics of Behavior Virtual Workshop (30 April 2020), organized by GJ Berman and GJ Stephens.  My sincere thanks to Gordon, Greg, and all my fellow panelists for this wonderful event, especially in these difficult times. Videos of the  lectures and discussion are available at https://www.youtube.com/watch?v=xSwWAgp2VdU.}

\author{William Bialek}

\affiliation{Joseph Henry Laboratories of Physics, and Lewis--Sigler Institute for Integrative Genomics, Princeton University, Princeton NJ 08544\\
Initiative for the Theoretical Sciences, The Graduate Center, City University of New York, 365 Fifth Ave, New York NY 10016}

\begin{abstract}
There is growing effort in the ``physics of behavior'' that aims at complete quantitative characterization of animal movements under more complex, naturalistic conditions.  One reaction to the resulting explosion of data is the search for low dimensional structure.   Here I try to define more clearly what we mean by the dimensionality of behavior, where observable behavior may consist either of  continuous  trajectories or sequences of discrete states. This discussion also serves to  isolate situations in which the dimensionality of behavior is effectively infinite.  I conclude with some more general perspectives about the importance of quantitative phenomenology.
\end{abstract}

\date{\today}

\maketitle

\section{Introduction}

Even large and complex animals have relatively small numbers of muscles or joints.  In some sense  the complexity of behavior is limited by this number of independent degrees of freedom.   Efforts to tame the complexity of brains and behavior have led to interest in a stronger notion, namely that the limited set of output degrees of freedom implies that the dimensionality of behavior is limited, and that correspondingly we should expect the dynamics of the neural networks that drive behavior also to be low--dimensional spaces.

There are many examples where we have direct evidence that motor behaviors are described by low--dimensional models, in organisms  from the worm {\em C elegans} to humans and nonhuman primates \cite{avella+bizzi_98,santello+al_98,sanger_00,osborne+al_05,stephens+al_08,stephens+al_11,ahamed+al_19}.   The argument that low dimensionality of behavior implies low dimensionality of neural activity has been made most explicitly for the case of {\em C elegans} \cite{kato+al_15,nichols+al_17},  but the search for low--dimensional manifolds in neural activity is  more widespread \cite{yu+al_09,gallego+al_17}.  Note that in the mammalian brain, with $\sim10^5$ neurons in one cortical column,  dimensionality could be reduced dramatically yet still be very large.

There are many reasons to be suspicious of the argument that a small number of behavioral degrees of freedom implies low dimensionality of behavior which in turn implies low dimensionality of neural activity.    There are, for example, $\sim 100$ muscles involved in human speech.  Does this mean that our linguistic behavior is $\sim 100$ dimensional?  Should we be searching for 100--dimensional dynamics in the patterns of neural activity that govern language production?  Should we be worried that there are $\sim 80$ muscles in the fruit fly thorax \cite{miller_50,lawrence_82}, which would mean that the potential dimensionalities of fly behavior and human language are not so different?  In fact even {\em C elegans} has 95 body wall muscles \cite{wormatlas}; the claim that the dynamics of worm behavior is low--dimensional rests on observations of the behavior itself, not on limits set by the anatomy.

Perhaps the comparison of flies and human language highlights the need for a more precise definition of  the ``dimensionality of behavior.'' This is made more urgent by the explosive growth of methods for more quantitative measurements of behavior \cite{stephens+al_08,branson+al_09,berman+al_14,wiltschko+al_15,mathis+al_18,pereira+al_19,datta+al_19,mathis+mathis_20}.  If these data can be reduced to low--dimensional descriptions, then we have achieved an enormous simplification, with practical consequences for further analysis.  We might also have theoretical  predictions about the dimensionality of behavior (either low or high), and then measuring dimensionality would provide decisive tests of these theories.
 
As a preface to the discussion, it should be remembered that the current wave  of quantitative approaches to the analysis of behavior integrates several very different intellectual traditions, including classical ethology, physiology, theoretical physics, engineering, and computer science.  Different ideas and methods will seem obvious or opaque to these very different communities.  In the interest of clarity I   take the risk of saying some things that will be well known to some audiences, and focus on what I hope are simple versions of general ideas.  
 
\section{Two examples}

To work toward a more precise definition, let's start with the case in which the behavior we observe is just a single function of time $x(t)$.   I will assume that this is a completely autonomous behavior,  and that across the time windows we consider the statistical structure of the behavior is stationary; both of these are common  but possibly unrealizable idealizations.   As is familiar from now classical literature on dynamical systems \cite{packard+al_80,abarbanel+al_93}, it is possible to tease out of this single time series a higher dimensional description of the underlying dynamics, so that the apparent dimensionality of the data is not a bound on the dimensionality of the dynamics.

On the other hand, suppose that a complete description of the observed behavior were given by 
\begin{equation}
\tau_c{{dx(t)}\over {dt}} = - x(t) + \eta(t),
\label{1D}
\end{equation}
where the $\eta(t)$ is white noise,
\begin{equation}
\langle  \eta(t) \eta(t')\rangle = {2\tau_c}\langle x^2\rangle \delta(t - t') .
\label{WN1}
\end{equation}
I think most people would agree that if this is a complete description of the dynamics, then the system really is one dimensional.  It is important that the noise source is white; non--white noise sources, which themselves are correlated over time, are equivalent to having hidden degrees of freedom that carry these correlations.  

The observable consequences of the   dynamics in Eqs (\ref{1D}, \ref{WN1}) are well known:  $x(t)$ will be a Gaussian stochastic process, with the two--point correlation function
\begin{equation}
C_1(\tau ) = \langle x(t) x(t+\tau ) \rangle = \langle x^2\rangle  e^{-|\tau|/\tau_c}.
\label{C1}
\end{equation}
We recall that for a Gaussian process, once we specify the two--point function there is nothing else to say about the system.
Importantly, we can turn this around:  if the observed behavior is a Gaussian stochastic process, and the correlations decay exponentially as in Eq (\ref{C1}), then Eqs (\ref{1D}, \ref{WN1}) are a complete description of the dynamics.  

Suppose the real dynamics involve not only the observable $x(t)$ but also an internal variable $y(t)$, 
\begin{equation}
\tau_c {d\over{dt}}
\begin{bmatrix}
x(t)  \\
y(t)
\end{bmatrix}
= 
-\begin{bmatrix}
1  & a  \\
a & 1 
\end{bmatrix}
\begin{bmatrix}
x(t)  \\
y(t)
\end{bmatrix}
+ 
\begin{bmatrix}
 \eta_1(t) \\
 \eta_2(t)
\end{bmatrix} ,
\label{2D1}
\end{equation}
where the driving noises  are white and independent,
\begin{equation}
\langle\eta_{\rm i} (t)\eta_{\rm j} (t')\rangle = 2 \tau_c \langle x^2 \rangle (1-a^2)  \delta_{\rm ij} \delta (t-t').
\end{equation}
Notice that since $y$ is hidden, the units of this variable are arbitrary, which allows us to have the strength of the noise driving each variable be the same without loss of generality, while the choice to give each variable the same correlation time is just for illustration, as is the symmetry of the dynamical matrix.  Looking at these equations, it seems easy to agree that the system is two dimensional.  Again, $x(t)$ again is Gaussian, but  the correlation function has two exponential decays,
\begin{widetext}
\begin{equation}
C_2(\tau ) = \langle x(t) x(t+\tau ) \rangle = {1\over 2} \langle x^2\rangle  (1-a^2) \left[ {1\over {1+a}} e^{-(1+a) |\tau|/\tau_c} + {1\over {1-a}} e^{-(1-a) |\tau|/\tau_c} \right] .
\label{C2}
\end{equation}
\end{widetext}

We see that a one dimensional system generates behavior with a correlation function that has one exponential decay, while a two dimensional system generates a correlation function with two exponential decays.  We would like to turn this around, and say that if we observe certain structure in the behavioral correlations, then we can infer the underlying dimensionality.

\section{Gaussian processes more generally}

Trying to analyze the structure of correlations by constructing explicit dynamical equations, as in Eqs (\ref{1D}) or (\ref{2D1}), may not be the best approach.  In particular, if there are hidden dimensions, then there is no preferred coordinate system in the space of unmeasured variables, and hence no unique form for the dynamical equations.  Let us instead focus on the probability distribution of trajectories $x(t)$.  For  Gaussian processes this has the form
\begin{eqnarray}
P[x(t)] &=& {1\over Z} e^{-S[x(t)]}\\
S[x(t)] &=& {1\over 2}\int dt \int dt' \, x(t) K(t-t') x(t'),
\label{K1}
\end{eqnarray}
where the integrals run over the interval of our observations, which should be long.  The kernel $K(\tau)$ is inverse to the correlation function,
\begin{equation}
\int dt'' \,K(t-t'') \langle x(t'' ) x(t')\rangle = \delta(t-t') .
\label{inv_def}
\end{equation}
We can divide the full trajectory $x(t)$ into the past, with $t\leq0$, and the future, with $t> 0$.  Schematically,  
\begin{eqnarray}
S[x(t)]  &=& {1\over 2} {\mathbf x}_{\rm past} \cdot K_{\rm pp} \cdot {\mathbf x}_{\rm past}+  {1\over 2} {\mathbf x}_{\rm fut} \cdot K_{\rm ff} \cdot {\mathbf x}_{\rm fut}  
\nonumber\\
&&\,\,\,\,\,\,\,\,\,\, + \, {\mathbf x}_{\rm past} \cdot K_{\rm pf} \cdot  {\mathbf x}_{\rm fut} ,
\end{eqnarray}
where $K_{\rm pf}$ couples the past and future.  More explicitly,
\begin{equation}
{\mathbf x}_{\rm past} \cdot K_{\rm pf} \cdot  {\mathbf x}_{\rm fut} = \int_0^\infty dt \int_0^\infty dt' \,x(-t) K(t+t') x(t') .
\label{Kpf_def}
\end{equation}
If $K_{\rm pf}$ is of finite rank, so that
\begin{equation}
K(t+t') = \sum_{{\rm n} =1}^D a_{\rm n} \phi_{\rm n}(t) \phi_{\rm n}(t'),
\label{K_D}
\end{equation}
then everything that we can predict about future behavior given knowledge of past behavior is captured by $D$ features,
\begin{eqnarray}
P[{\mathbf x}_{\rm fut} | {\mathbf x}_{\rm past}] &=& P[{\mathbf x}_{\rm fut} | \{F_{\rm n}\}]
\label{suffstat}\\
F_{\rm n} &=& \int_0^\infty dt \,\phi_{\rm n}(t) x(-t) .
\end{eqnarray}

Equation (\ref{suffstat}) is telling us that the features $\{F_{\rm n}\}$ provide ``sufficient statistics'' for making predictions.   We recall that in a dynamical system with $D$ variables, 
\begin{equation}
{{dy_{\rm i}}\over{dt}} = g_{\rm i}(\{y_{\rm j}\}) + \eta_{\rm i}(t) ,\,\,\,\,\, {\rm i} = 1,\, 2,\, \cdots ,\, D,
\label{D-dim-dyn}
\end{equation}
predicting the future ($t>0$) requires specifying   $D$ initial conditions (at $t=0$).  In this precise sense, the number of variables that we need to achieve maximum predictive power {\em is} the dimensionality of the dynamical system.  To complete the argument, we need to show that $K_{\rm pf}$ has finite rank when  correlations decay as a finite combination of exponentials; see Appendix \ref{app:K}.

In the case of Gaussian stochastic processes we thus arrive at a recipe for defining the dimensionality of the underlying dynamics.  We estimate the correlation function, take its inverse to find  the kernel, and isolate the part of this kernel which couples past and future.  If this past--future kernel is of finite rank, then we can identify this rank with the dimensionality of the system.

This discussion still refers only to Gaussian processes, but we see that the search for low--dimensional descriptions could fail, qualitatively.  It is possible that the past--future kernel $K_{\rm pf}$ is {\em not} of finite rank; more generally  if we analyze signals in a window of size $T$ then the rank can grow with $T$.    This happens, for example, if  behavioral correlations decay as a power of time,
\begin{equation}
\langle x(t) x(t')\rangle = {{t_0^\alpha}\over{t_0^\alpha + |t-t'|^\alpha}} .
\label{power1}
\end{equation}
Under these conditions the system is infinite dimensional.

\section{Beyond  Gaussians}

What emerges from the analysis of Gaussian stochastic processes is that dimensionality can be measured through the problem of prediction.  Let us see how we can make this more general, beyond the Gaussian case.

Let us break a very long observation of time series into many examples of a time window $-T < t < T$.  Within each window,  the trajectory $x(t < 0)$ defines the past ${\mathbf x}_{\rm past}$, and $x(t>0)$ defines the future ${\mathbf x}_{\rm fut}$.  Across a large ensemble of these windows we can define the joint probability distribution $P_T({\mathbf x}_{\rm past}, {\mathbf x}_{\rm fut})$.  To characterize the {\em possibility} of making predictions we can measure the mutual information between past and future, or the ``predictive information'' \cite{bialek+al_01},
\begin{widetext}
\begin{equation}
I_{\rm pred}(T) \equiv I({\mathbf x}_{\rm past}; {\mathbf x}_{\rm fut}) = \sum_{{\mathbf x}_{\rm past}}\sum_{ {\mathbf x}_{\rm fut}} 
P_T({\mathbf x}_{\rm past}, {\mathbf x}_{\rm fut}) 
\log\left[
{
{P_T({\mathbf x}_{\rm past}, {\mathbf x}_{\rm fut})}
\over
{P_T({\mathbf x}_{\rm past}) P_T( {\mathbf x}_{\rm fut})}
}
\right].
\end{equation}
\end{widetext}

The predictive information can have very different qualitative behaviors as $T$ becomes large \cite{bialek+al_01}.\footnote{If we observe a continuous variable $x(t)$ in continuous  time, then smoothness across $t=0$ generates a formal divergence in the mutual information between past and future. Modern analyses of behavior typically begin with video data, with time in discrete frames, evading this problem.  Alternatively, if measurements on $x(t)$ include a small amount of white noise, then the predictive information becomes finite even without discrete time steps. Thanks to A Frishman for emphasizing the need for care here.}
   For a time series that can be captured by a finite state Markov process, or more generally described by a finite correlation time, then $I_{\rm pred}(T)$ is finite as $T\rightarrow \infty$.  On the other hand, for Gaussian processes with correlation functions that decay as a power, as in Eq (\ref{power1}), the predictive information diverges logarithmically, $I_{\rm pred}(T \rightarrow \infty) \propto \log T$.

In the example of a dynamical system with $D$ variables, as in Eq (\ref{D-dim-dyn}), all the predictive power available will be realized if we can specify $D$ numbers, which are the initial conditions for integrating the differential equations.  Thus we consider mappings of the  past  into $d$ features,
\begin{equation}
{\cal M}_d : {\mathbf x}_{\rm past} \rightarrow \{F_\mu\}, \,\,\, \mu = 1,\, 2,\, \cdots,\,d .
\end{equation}
For any choice of features we can compute how much predictive information has been captured, and then we can maximize over the mapping, resulting in
\begin{equation}
I_{\rm pred}(T; d ) = \max_{{\cal M}_d } I(\{F_\mu\}; {\mathbf x}_{\rm fut}),
\end{equation}
which is the maximum predictive information we can capture with $d$ features in windows of duration $T$.

If the system truly is $D$ dimensional, then  $D$ features of the past are sufficient to capture all of the available predictive information.  This means that a plot of $I_{\rm pred}(T; d )$ vs $d$ will saturate.  To be precise we are interested  in what happens at large $T$, so we can define 
\begin{equation}
\lim_{T\rightarrow\infty} {{I_{\rm pred}(T; d )}\over{I_{\rm pred}(T)}} = f(d).
\end{equation}
If we find that $f(d\geq D )= 1$, then we can conclude that the behavior has dimensionality $D$.

In words, the dimensionality of behavior is the minimum number of features of the past needed to make maximally informative predictions about the future, and to be precise the past should be taken to be of long duration.  While very general, it should be admitted that this definition is much more complex than the analysis of correlation functions that works in the Gaussian case.  To use this definition we have to search over all possible mappings ${\cal M}_d$ of long past trajectories into $d$--dimensional feature spaces, and we have to estimate the mutual information between these $d$ variables and some representation of the future trajectory.  Both of these steps are challenging.

\section{Discrete states}

In many cases it is natural to  describe animal behavior as moving through a sequence of discrete states.  We do this, for example, when we transcribe human speech to text, and  when we describe a bacterium as running or tumbling \cite{berg+brown_72}.  This identification of discrete states is not just an arbitrary quantization of continuous motor outputs, nor  should it be a qualitative judgement by human observers. Discrete states should correspond to distinguishable clusters, or resolvable peaks in the distribution over the natural continuous variables, and the dynamics should consist of movements in the neighborhood of one peak that are punctuated by relatively rapid jumps to another peak (e.g., Ref \cite{berman+al_14}).  A ``mechanism'' for such discreteness is the existence of multiple dynamical attractors, with jumps driven by noise (e.g., Refs \cite{stephens+al_08,stephens+al_11}).

When behavioral states are discrete, how do we define dimensionality?  Once again it is useful to think about the simplest case, where there are just two behavioral states---perhaps ``doing something'' and ``doing nothing''---and time is marked by discrete ticks of a clock.  We can represent the two states at each time $t$ by an Ising variable $\sigma_t = \pm 1$.   If the sequence of behavioral states were Markovian, then $\sigma_t$ depends only on $\sigma_{t-1}$, and because $\sigma^2 = 1$ the only possible stationary probability distribution for the sequences $\sigma_1,\, \sigma_2, \, \cdots ,\, \sigma_T$ is
\begin{equation}
P\left( \{\sigma_t\}\right) = {1\over Z}\exp\left[ h\sum_t \sigma_t + J\sum_t \sigma_{t-1}\sigma_t \right] ,
\end{equation}
 which is the one--dimensional Ising model with nearest neighbor interactions.  Importantly, if we measure the correlations of the fluctuations in behavioral state around its mean,
 \begin{equation}
C(t-t') \equiv \langle \left( \sigma_t - \langle\sigma\rangle\right) \left( \sigma_{t'} - \langle\sigma\rangle\right)\rangle ,
\end{equation}
we find that these correlations decay exponentially, 
\begin{equation}
C(t-t') = C(0) e^{-|t-t'|/\tau_c},
\end{equation}
where we can express $\tau_c$ in terms of $h$ and $J$ \cite{1DIsing}.  This reminds us of the exponential decays  in the continuous case with Gaussian fluctuations, where they provide a signature of low dimensionality.  

Suppose that we have only two states, but observe correlations that do not decay as a single exponential.  Then the probability distribution $P\left( \{\sigma_t\}\right)$ must have terms that describe explicit dependences of $\sigma_t$ on $\sigma_{t'}$ with $t-t' > 1$.  This can  be true only if there are some hidden states or variables that carry memory across the temporal gap $t-t'$.  A sensible definition for the dimensionality of behavior then refers to these internal variables.

Imagine that we observe the mean of the behavioral variable, $\langle\sigma\rangle$, and the correlation function $C(t-t')$.  What can we say about the probability distribution $P\left( \{\sigma_t\}\right)$?  There are infinitely many models that are consistent with measurements of just the (two--point) correlations, but there is one that stands out as having minimal structure required to match these observations \cite{jaynes_57}.  Said another way, there is a unique model that predicts the observed correlations but otherwise generates behavioral sequences that are as random as possible.  This minimally structured model is the one that has the largest possible entropy, 
and it has the form
\begin{equation}
P\left( \{\sigma_t\}\right) = {1\over Z}\exp\left[ h\sum_t \sigma_t + {1\over 2}\sum_{t,t'} J(t-t') \sigma_{t}\sigma_{t'} \right] ,
\label{maxent1}
\end{equation}
where the parameter $h$ must be adjusted so that the model predicts the observed mean behavior $\langle\sigma\rangle$, and the function $J(t-t')$ must be adjusted so that the model predicts the observed correlation function $C(t-t')$.  

Maximum entropy models have a long history, and a deep connection to statistical mechanics \cite{jaynes_57}.  As applied to temporal sequences, the maximum entropy models sometimes are referred to as maximum caliber \cite{dixit+al_18}.  The Boltzmann distribution, which describes a system in thermal equilibrium, can be derived as the maximum entropy distribution over microscopic states of the the system that is consistent with its mean energy, and this sometimes leads to the impression that maximum entropy models only describe equilibrium systems,  but this isn't correct.  In this discussion we are explicitly using the maximum entropy idea to describe distributions of sequences or trajectories, not distributions over states as with the Boltzmann distribution. In the case of just two states, if we only constrain the two--point correlations $C(t-t')$ we cannot distinguish the arrow of time, but as soon as we have more states, or constrain higher--order correlations, the maximum entropy model can  break time--reversal invariance or detailed balance.  For biological systems there has been interest in the use of maximum entropy methods to describe amino acid sequence variation in protein families \cite{bialek+ranganathan_07,weigt+al_09,marks+al_11}, patterns of electrical activity in populations of neurons \cite{schneidman+al_06,shlens+al_09,sdme,tkacik+al_14,meshulam+al_17},  velocity fluctuations in flocks of birds \cite{bialek+al_12,bialek+al_14}, and more.  There have been more limited attempts to use these ideas in describing temporal sequences, either in neural populations \cite{mora+al_15} or flocks \cite{cavagna+al_14a,mora+al_16,ferrretti+al_20}.

The maximum entropy model in Eq (\ref{maxent1})  can be rewritten exactly as a model in which the behavioral state at time $t$ depends only on some internal variable $x(t)$:
\begin{eqnarray}
P\left( \{\sigma_t\}\right) &=& \int Dx \,P[x(t)] \prod_t P(\sigma_t | x(t) +h ), \\
P(\sigma | x +h ) &=& {{\exp\left[\sigma\cdot\left( x + h\right) \right]}\over{2 \cosh (x+h)}} ,
\end{eqnarray}
and the distribution of the internal variable is
\begin{eqnarray}
P[x(t)] &=& {1\over{Z'}} e^{-S'[x(t)]}\\
S'[x(t)] &=&   {1\over 2}\sum_{t,t'} x(t) K(t-t') x(t')\nonumber\\
&&\,\,\,\,\,\,\,\,\,\, - \sum_t \ln\cosh \left( x(t) + h\right)  ,
\label{PX}
\end{eqnarray}
where $K(t)$ is the matrix inverse of the function $J(t)$,
\begin{equation}
\sum_{t''} K(t - t'') J(t'' - t') = \delta_{tt'} .
\end{equation}
Notice that since $J(0)$ multiplies $\sigma_t \sigma_t = 1$, its value can't change the observable statistics of behavior, so we have some freedom in writing the model this way.\footnote{Models where the observed degrees of freedom depend on hidden or latent variables, but not directly on one another, are sometimes set in opposition to statistical physics models, where it is more natural to think about direct interactions.  But this example shows that these pictures can be mathematically equivalent; see also the Supplementary material of Ref \cite{tkacik+al_15}.}

Starting from discrete binary states we thus are led back to an underlying continuous variable, and we can carry over our definitions of dimensionality.   Although $x(t)$ is not  Gaussian, the only coupling of past and future is through a kernel $K(t)$, just as in the Gaussian case, as we see by comparing Eqs (\ref{K1}) and (\ref{PX}).  This kernel is not the inverse of the observed behavioral correlations, but of the effective interactions between states at different times, $J(\tau)$.  But, importantly, we are considering quantities that are {\em determined} by the correlation function and hence the problem is conceptually similar to the Gaussian case:  we analyze the correlations to derive a kernel, and the dimensionality of behavior is the rank of this kernel.  The maximum entropy model plays a useful role because it is the least structured model consistent with the observed correlations.

If $x(t)$ is one--dimensional,   then the interactions $J(t) \sim J_0 e^{-|t|/\tau}$.  The correlations $C(t)$ are predicted to decay exponentially at large $|t|$, but by the time this limit is reached the correlations may be so weak that this is hard to measure convincingly.  At the more accessible intermediate times the behavior of $C(t)$ can be complicated even though the underlying dynamics are one--dimensional.  At the opposite extreme, if $x(t)$ has effectively infinite dimensionality, then we can have $J(t)\sim J_0|t|^{-\alpha}$, as in Eq (\ref{power1}).  Ising models with such power--law interactions are the subject of a large literature in statistical physics; the richest behaviors are at $\alpha = 2$,  where results  presaged major developments in the renormalization group and topological phase transitions \cite{ruelle_68,dyson_69,yuval+anderson_70,yuval+al_70}.   It would be fascinating if these models emerged as  effective descriptions of strongly non--Markovian sequences in animal behavior.

\section{Perspectives}

The explosion of quantitative data on animal behavior is exciting in part because these essentially macroscopic behaviors---rather than their microscopic mechanisms---are what first strike us as being interesting about living systems.  Behaviors have been selected by evolution for their utility, and as we observe them it is difficult not to think of them as purposeful or intelligent.   Understanding the phenomena of life means explaining how these behaviors arise, ultimately from interactions among molecules that obey the same laws of physics as in inanimate (and unintelligent) matter.  But what is it, exactly, that we are trying to explain?

If we want to explain why we look like our parents, a qualitative answer is that we carry copies of their DNA.  But if we want to understand the reliability with which traits are passed from generation to generation,\footnote{A poetic formulation of this problem is Schr\"odinger's description of the Hapsburger lippe,   passed for centuries through the imperial family of Austria  \cite{schrodinger_44}.} then talking about DNA structure is not enough---the energy differences between correct and incorrect base pairing are not sufficient to explain the reliability of molecular copying if the reactions are allowed to come to thermal equilibrium, and this problem arises not just in DNA replication but in every step of molecular information transmission.  Cells achieve reliability by holding these reactions away from equilibrium, allowing for proofreading or error--correction \cite{hopfield_74,ninio_75}.  In the absence of proofreading, the majority of proteins would contain at least one incorrect amino acid, and roughly 10\% of our genes would be different from those carried by either parent;  these error rates are orders of magnitude larger than  observed.  These quantitative differences are so large that life without proofreading would be qualitatively different.\footnote{In retroviruses, including HIV, reproduction includes a step of reverse transcription, which occurs without proofreading.  The dramatically accelerated pace of evolution in these viruses gives a glimpse of how different life would be if the transmission of genetic information depended on base pairing alone.}

The example of proofreading highlights the importance of starting with a  quantitative  characterization of the phenomena we are trying to explain.  In an era of highly mechanistic biology, this emphasis on phenomenological description  may seem odd.  But quantitative phenomenology has been  foundational, certainly in physics and also in the mainstream of biology.  Mendel's genetics was a phenomenological description of the patterns of inheritance, and the realization that genes are arranged linearly along chromosomes came from a more refined quantitative analysis of these same patterns \cite{sturtevant_13}.  The work of Hodgkin and Huxley led to our modern understanding of electrical activity  in terms of ion channel dynamics, but explicitly eschewed mechanistic claims in favor of phenomenology  \cite{hodgkin+huxley_52}.  The idea that transmission  across a synapse depends on transmitter molecules packaged into vesicles emerged from the quantitative analysis of voltage fluctuations at the neuromuscular junction \cite{fatt+katz_52}.

Even when we are searching for microscopic mechanisms, it is not anachronistic to explore macroscopic descriptions.    Time and again, the scientific community has leaned on phenomenology to imagine the underlying mechanisms, often taking literally the individual terms in a mathematical description as representing the actual microscopic elements for which we should be searching, whether these are genes, ion channels, synaptic vesicles, or quarks \cite{zweig_80,zweig_10,zweig_14}.  What is anachronistic, in the literal sense of the word, is to believe that microscopic mechanisms were discovered by direct microscopic observations without guidance from phenomenology on a larger scale.

The idea that quantitative phenomenology would provide a foundation for understanding brains and minds took hold very early in the modern era.  In the second half of the nineteenth century, many people were trying to turn observations on seeing and hearing into quantitative experiments, creating a subject that would come to be called psychophysics \cite{green+swets_66}. By $\sim$1910, these experiments were sufficiently well developed that Lorentz could look at data on the ``minimum visible'' and suggest that the retina is capable of counting single photons \cite{bouman_61}, and Rayleigh could identify the conflict between our ability to localize low--frequency sounds and the conventional wisdom that we are ``phase deaf'' \cite{rayleigh_07}.  Both of these essentially theoretical observations, grounded in quantitative descriptions of human behavior, would drive experimental efforts that unfolded over more than fifty years.

Also $\sim$1910, von Frisch was doing psychophysics experiments  to demonstrate bees could, in fact, discriminate among the beautiful colors of the flowers that they pollinate \cite{frisch_74}.  But he took these experiments in a very different direction, focusing not on the discrete choices made by individual bees, but on how these individuals communicated their sensory experiences to other residents of the hive.  This work led to the discovery of the ``dance language'' of bees.   While von Frisch often used simplified stimuli, and counted whether bees arrived at a destination or not, it was crucial that intermediate behaviors---the dance---were unconstrained and fully natural.  

What grew out of the work by von Frisch and others  was the field of ethology \cite{gould_82},  which emphasizes the richness of behavior in its natural context, the context in which it was selected for by evolution. Because ethologists wrestle with complex behaviors, they often resort to verbal description.   In contrast, psychophysicists  focus on situations in which subjects are constrained  to  a small number of discrete alternative behaviors, so it is natural to give a quantitative description by estimating the probabilities of different choices under various conditions. 

The emergence of a quantitative language for the analysis of psychophysical experiments was aided by the focus on constrained behaviors, but was not an automatic consequence of this focus.  For photon counting in vision, the underlying physics suggests how the probability of seeing vs.~not seeing will depend on light intensity \cite{hecht+al_42}, but the observation that human observers behave as predicted points to profound facts about  the underlying mechanisms \cite{bialek_12}.  During World War II, attempts to formalize the problems of radar operators and communication with pilots led to a more general view of the choices among discrete alternative behaviors as being discriminations among signals in a background of noise \cite{lawson+uhlenbeck_50}. In the 1950s and 60s this view was exported to experimental psychology and developed further into the modern form of signal detection theory \cite{green+swets_66}.   Much of this now seems like an exercise in probability and statistics, something obviously correct, but the early literature records considerable skepticism about whether this (or perhaps any) mathematization of human behavior would succeed.

Much has been learned through both the ethological and the psychophysical approaches, yet it is easy for advocates of the two approaches to talk past one another.
Still,  it does not seem unfair to note that the traditional ethological approach is missing the drive for quantification, while the traditional psychophysical approach achieves  quantitative sophistication by excluding much of what impresses us about  behavior.  The challenge is not to find what each tradition is lacking, but to find a way of combining the best from both, and this brings us back to the questions asked at the outset.

A quantitative characterization of naturalistic behaviors requires that we  attach comparable numbers to very different kinds of time series.  Dimensionality is a candidate for this sort of characterization. When we do psychophysics, we characterize behaviors with numbers that are meaningfully comparable across situations and across species.  To give but one example, we can discuss the accumulation of evidence for decisions that humans and non--human primates make based on visual inputs, but we can use the same mathematical language to discuss decisions made by rodent based on auditory inputs \cite{brunton+al_13}.   Perhaps the dimensionality of behavior will provide part of the needed unifying mathematical language for more natural behaviors.

\begin{acknowledgments}
Thanks to V Alba, GJ Berman, X Chen, A Frishman, K Krishnamurthy, AM Leifer, CW Lynn, SE Palmer, JW Shaevitz, and GJ Stephens,  for many helpful discussions.  This work was supported in part by the National Science Foundation through the Center for the Physics of Biological Function (PHY--1734030) and Grant PHY--1607612, and by the National Institutes of Health (NS104889).
\end{acknowledgments}

\appendix

\section{Past-future kernels, explicitly}
\label{app:K}

We are interested in the behavior of the kernel $K$ when the correlation function $C$ is a sum of exponentials.  As noted above, we need to be a little careful to make this problem well posed. If we monitor a continuous variable in continuous time, then continuity leads to an infinite mutual information between $x(t^-)$ and $x(t^+)$.  We can solve this either by assuming that observation are made at discrete ticks of a clock (as in video recordings), or by assuming that observations are made in a background of white noise.  Here I will take the second approach.

The statement that the correlation function is a sum of exponentials, but measurements are in a background of white noise, means that the observed correlation function 
\begin{equation}
\langle x(t) x(0)\rangle \equiv C(t ) = \sum_{\mu =1}^MA_\mu e^{-|t|/\tau_\mu} + {\cal N}\delta(t),
\label{A1}
\end{equation}
where ${\cal N}$ is the strength of the noise.  We want to construct the kernel $K(t)$ that is the operator inverse to $C$, as in Eq (\ref{inv_def}).  We recall that this can be done by passing to Fourier space:
\begin{eqnarray}
G(\omega ) &=& \int dt \, e^{+i\omega t} C(t)\\
K(t) &=& \int {{d\omega}\over{2\pi}} e^{-i\omega t} {1\over{G(\omega )}} .
\label{KfromG}
\end{eqnarray}
From Eq (\ref{A1}) we can see that
\begin{eqnarray}
G(\omega ) &=& \int dt \, e^{+i\omega t} \left[\sum_{\mu =1}^M A_\mu e^{-|t|/\tau_\mu} + {\cal N}\delta(t)\right]\\
&=& \sum_{\mu =1}^M {{2A_\mu\tau_\mu}\over{1+(\omega\tau_\mu)^2}} + {\cal N} .
\end{eqnarray}
Then to find $K(t)$ we invert and transform back, being careful to isolate the contribution of the white noise term:
\begin{eqnarray}
K(t) &=& \int {{d\omega}\over{2\pi}} e^{-i\omega t} \left[\sum_{\mu =1}^M {{2A_\mu\tau_\mu}\over{1+(\omega\tau_\mu)^2}} + {\cal N}\right]^{-1}\\
&=&\int {{d\omega}\over{2\pi}} e^{-i\omega t} \left[ {1\over{\cal N}} - {{P_{M-1}(\omega^2)}\over{P_M(\omega^2)}}\right],
\label{int1}
\end{eqnarray}
where
\begin{equation}
P_{M-1}(\omega^2) =  \sum_{\mu=1}^M 2A_\mu\tau_\mu\prod_{\nu\neq\mu} [1 + (\omega\tau_\nu)^2]
\end{equation}
is a $M-1$st order polynomial in $\omega^2$ and
\begin{equation}
P_{M}(\omega^2) =  {\cal N}\left({\cal N}\prod_{\mu =1}^M [1+(\omega\tau_\mu)^2] +P_{M-1}(\omega^2)\right)
\end{equation}
is a $M$th order polynomial in $\omega^2$.  Note that both polynomials have all real and positive coefficients.

We notice that $P_{M-1}(\omega^2) /P_{M}(\omega^2)$ vanishes at large $|\omega|$, and $e^{-i\omega t}$ vanishes for values of $\omega$ with large negative (positive) imaginary part if $t>0$ ($t<0$).   This means that we can do the integral over $\omega$ in Eq (\ref{int1}) by closing a contour in the complex plane.  Then we can use the fact that  
\begin{equation}
P_M (\omega^2) = B \prod_{{\rm n} = 1}^M (\omega^2 - \omega_{\rm n}^2),
\end{equation}
where $B$ is a constant and the $\{\omega_{\rm n}^2\}$ are the roots of the polynomial.  The simplest case is where all $\omega_{\rm n}^2$ are real, in which case they must be negative and we can write $\omega_{\rm n} =  -i\lambda_{\rm n}$, with $\lambda_{\rm n} > 0$.  Then for $t>0$ we close the contour in the lower half plane, which picks out the poles at $\omega = \omega_{\rm n}$, while for $t<0$ we close the contour in the upper half plane, which picks out the poles at $\omega = -\omega_{\rm n}$.  The result is that
\begin{eqnarray}
K(t ) &=& {1\over {\cal N}} \delta(t) - \int {{d\omega}\over{2\pi}} e^{-i\omega t}{{P_{M-1}(\omega^2)}\over{P_M(\omega^2)}} \\
&=&{1\over {\cal N}} \delta(t)- {1\over B} \sum_{\rm n} {{P_{M-1}(\omega^2 = -\lambda_{\rm n}^2)}\over {2\lambda_{\rm n}\prod_{{\rm m}\neq{\rm n}}(\lambda_{\rm m}^2 - \lambda_{\rm n}^2) }}e^{-\lambda_{\rm n}|t|}.\nonumber\\
&&
\end{eqnarray}

If we look back at the derivation of Eq (\ref{Kpf_def}), we can see that a delta function term in $K(t)$ does not contribute to coupling past and future.  Thus $K(t>0)$ collapses into the form of Eq (\ref{K_D}):
\begin{eqnarray}
K(t+t') &=&\sum_{{\rm n}=1}^M a_{\rm n}\phi_{\rm n}(t) \phi_{\rm n}(t') \\
a_{\rm n} &=& {1\over B}{{P_{M-1}(\omega^2 = -\lambda_{\rm n}^2)}\over {2\lambda_{\rm n}\prod_{{\rm m}\neq{\rm n}}(\lambda_{\rm m}^2 - \lambda_{\rm n}^2) }}\\
\phi_{\rm n}(t) &=& e^{-\lambda_{\rm n} t},
\end{eqnarray}
and the dimensionality $D = M$, as we hoped: if the observed behavioral variable is Gaussian, and the correlation function can be written as the sum of $M$ exponentials, then the system has underlying dimensionality $D=M$.

It is useful to work out the case $M=1$.  Then we have
\begin{equation}
\langle x(t) x(0)\rangle \equiv C(t ) = A e^{-|t|/\tau_c} + {\cal N}\delta(t),
\end{equation}
and after some algebra we find
\begin{eqnarray}
K(t) &=& a_1 e^{-\lambda_1 |t|}\label{K1A}\\
\lambda_1 &=& {1\over{\tau_c}}\sqrt{1 + 2A\tau_c/{\cal N}} .
\end{eqnarray}
It is useful to think more explicitly about the fact that we have embedded a correlated signal in the background of white noise, so we can write
\begin{eqnarray}
x(t) &=& y(t) + \eta(t)\\
\langle y(t) y(t')\rangle &=& A e^{-|t-t'|/\tau_c} \\
\langle \eta (t) \eta(t')\rangle &=& {\cal N}\delta(t-t').
\end{eqnarray}
Only $y(t)$ is predictable,  the best predictions would be based on knowledge of $y(t=0)$.  One can then show that the best estimate of this quantity given observations on the noisy $x(t)$ is
\begin{equation}
y_{\rm est}(0) = \int_0^\infty dt\, K(t) x(-t),
\end{equation}
with the same $K(t)$ as in Eq (\ref{K1A}).  Thus, asking for the optimal prediction is the same as asking for the optimal separation of the predictable signal from the unpredictable noise \cite{bialek+al_07}.


\begin{thebibliography}{99}
%
\bibitem{avella+bizzi_98}
A d’Avella and E Bizzi,  Low dimensionality of surpaspinally induced force fields. {\em Proc Natl Acad Sci (USA)} {\bf 95,} 7711--7714 (1998).
%
\bibitem{santello+al_98}
M Santello, M Flanders, and JF Soechting,  Postural hand strategies for tool use. {\em J Neurosci} {\bf 18,} 10105--10115 (1998).
%
\bibitem{sanger_00}
TD Sanger, Human arm movements described by a low--dimensional superposition of principal components. {\em J Neurosci} {\bf 20,} 1066--1072 (2000).
%
\bibitem{osborne+al_05}
 LC Osborne, SG Lisberger, and W Bialek,   A sensory source for motor variation.  {\em Nature} {\bf 437,} 412--416 (2005).
%
\bibitem{stephens+al_08}
GJ Stephens, B Johnson--Kerner, W Bialek, and WS Ryu,    Dimensionality and dynamics in the behavior of {\em C. elegans}.    {\em PLoS Comput Biol} {\bf 4,} e1000028 (2008)
%
\bibitem{stephens+al_11}
GJ Stephens, MB de Mesquita, WS Ryu, and W Bialek,  Emergence of long timescales and stereotyped behaviors in {\em Caenorhabditis elegans}.   {\em Proc Natl Acad Sci (USA)} {\bf 108,} 7286--7289 (2011).
 %
 \bibitem{ahamed+al_19}
T Ahamed, AC Costa, and GJ Stephens, Capturing the continuous complexity of behavior in {\em C elegans}. arXiv:1911.10559 [q--bio.NC] (2019).
%
\bibitem{kato+al_15}
S Kato, HS Kaplan, T Schr\"odel, S Skora, TH Lindsay, E Yemini, S Lockery, and M Zimmer, Global brain dynamics embed the motor command sequence of {\em Caenorhabditis elegans}. {\em Cell} {\bf 163,} 656--669 (2015).
%
\bibitem{nichols+al_17}
ALA Nichols, T Eichler, R Latham, and M Zimmer, A global brain state underlies {\em C. elegans} sleep behavior. {\em Science} {\bf 356,}  eaam6851 (2017).
%
\bibitem{yu+al_09}
BM Yu, JP Cunningham, G Santhanam, SI Ryu, KV Shenoy,  and M Sahani, Gaussian--process factor analysis for low--dimensional single--trial analysis of neural population activity. {\em J Neurophysiol} {\bf 102,} 614--635 (2009).  
%
\bibitem{gallego+al_17}
JA Gallego, MG Perich, LE Miller, and SA Solla, Neural manifolds for the control of movement. {\em Neuron} {\bf 94,} 978--984 (2017).
%
\bibitem{miller_50}
A Miller, The anatomy and histology of the imago of {\em Drosophila melanogaster}. In {\em The Biology of Drosophila}, M Demerec, ed, pp 420--534 (Wiley, New York, 1950).  
%
\bibitem{lawrence_82}
PA Lawrence, Cell lineage of the thoracic muscles of {\em Drosophila}.  {\em Cell} {\bf 29,} 493--5050 (1982).
%
\bibitem{wormatlas}
ZF Altun  and DH Hall,   Muscle system, introduction. In {\em WormAtlas}, ZF Altun, L Herndon, CA Wolkow,  C Crocker, R Lints, and DH Hall,  eds,  doi:10.3908/wormatlas.1.6 (2009).
%
\bibitem{branson+al_09} 
K Branson, AA Robie, J Bender, P Perona, and MH Dickinson, High--throughput ethomics in large groups of {\em Drosophila}. {\em Nature methods} {\bf 6,} 451--457 (2009).
%
\bibitem{berman+al_14}
GJ Berman,  DM Choi, W Bialek, and JW Shaevitz, Mapping the stereotyped behaviour of freely moving fruit flies.  {\em J R Soc Interface} {\bf 11,} 20146072 (2014).
%
\bibitem{wiltschko+al_15}
AB Wiltschko, MJ Johnson, G Iurilli, RE Peterson, JM Katon, SL Pashkovski, VE Abraira, RP Adams, and SR Datta,  Mapping sub--second structure in mouse behavior.  {\em Neuron} {\bf 88,} 1--15 (2015).
%
\bibitem{mathis+al_18}
A Mathis, P Mamidanna, KM Cury, T Abe, VN Murthy, MW Mathis, and  M Bethge, DeepLabCut: Markerless pose estimation of user-defined body parts with deep learning.  {\em Nature Neurosci} {\bf 21,} 1281--1289 (2018).
%
\bibitem{pereira+al_19}
T Pereira, D Aldarondo, L Willmore, M Kislin, SS Wang, M Murthy, and JW Shaevitz, Fast animal pose estimation using deep neural networks. {\em Nature Methods} {\bf 16,} 117--125 (2019).
%
\bibitem{datta+al_19}
SR Datta, DJ Anderson, K Branson, P Perona, and A Leifer,  Computational neuroethology: A call to action. {\em Neuron} {\bf 104,} 11--24 (2019).
%
\bibitem{mathis+mathis_20}
MW Mathis and A Mathis, Deep learning tools for the measurement of animal behavior in neuroscience.
{\em Current Opin Neuro} {\bf 60,} 1--11 (2020).
%
\bibitem{packard+al_80}
NH Packard, JP Crutchfield, JD Farmer, and RS Shaw, Geometry from a time series. {\em Phys Rev Lett} {\bf 45,} 712--716 (1980).
%
\bibitem{abarbanel+al_93}
HDI Abarbanel, R Brown, JJ Sidorowich, and LS Tsimring, The analysis of observed chaotic data in physical systems.  {\em Rev Mod Phys} {\bf 65,} 1331--1392 (1993).
%
\bibitem{bialek+al_01}
W Bialek, I Nemenman, and N Tishby,  Predictability, complexity and learning. {\em Neural Comp} {\bf 13,} 2409--2463 (2001).
%
\bibitem{berg+brown_72}
HC Berg and DA Brown, Chemotaxis in {\em Escherichia coli} analysed by three--dimensional tracking. {\em Nature} {\bf 239,} 500--504 (1972).
%
\bibitem{1DIsing}
CJ Thompson, {\em Mathematical Statistical Mechanics} (Princeton University Press, Princeton NJ, 1972).
%
\bibitem{jaynes_57} 
ET Jaynes, Information theory and statistical mechanics. {\em Phys Rev} \textbf{106,} 620--630 (1957). 
%
\bibitem{dixit+al_18}
PD Dixit, J Wagoner,   C Weistuch,   S Press\'e, K Ghosh, and KA Dill, Perspective: Maximum caliber is a general variational principle for dynamical systems. {\em J Chem Phys} {\bf 148,} 010901 (2018).
%
\bibitem{bialek+ranganathan_07} 
W Bialek and R Ranganathan, Rediscovering the power of pairwise interactions. arXiv.org:0712.4397 [q--bio.QM] (2007).
%
\bibitem{weigt+al_09}
 M Weigt, RA White, H Szurmant, JA Hoch, and T Hwa, Identification of direct residue contacts in protein--protein interaction by message passing.  {\em Proc Natl Acad Sci USA} {\bf 106,} 67--72  (2009).
%
 \bibitem{marks+al_11}
 DS Marks, LJ Colwell, R Sheridan, TA Hopf, A Pagnani, R Zecchina, and C Sander, Protein 3D structure computed from evolutionary sequence variation.  {\em PLoS One} {\bf 6,} e28766  (2011) .
 %
\bibitem{schneidman+al_06}
E Schneidman, MJ Berry II, R Segev, and  W Bialek, Weak pairwise correlations imply strongly correlated network states in a neural population. {\em Nature} {\bf 440,} 1007--1012  (2006).
%
\bibitem{shlens+al_09}
J Shlens, GD Field, JL Gaulthier, M Greschner, A Sher, AM Litke, and EJ Chichilnisky, The structure of large--scale synchronized firing in primate retina.  {\em J Neurosci} {\bf 29,} 5022--5031 (2009).
%
\bibitem{sdme}
E Granot-Atedgi, G Tka\v{c}ik, R Segev, and E Schneidman, Stimulus-dependent maximum entropy models of neural population codes. \emph{PLoS Comput Biol} {\bf 9,} e1002922 (2013). 
%
\bibitem{tkacik+al_14}
G Tka\v{c}ik, O Marre, D Amodei, E Schneidman, W Bialek, and MJ Berry II, Searching for collective behavior in a large network of sensory neurons.  {\em PLoS Comput Biol} {\bf 10,} e1003408 (2014).
%
\bibitem{meshulam+al_17}
 L Meshulam, JL Gauthier, CD Brody, DW Tank, and W Bialek,  Collective behavior of place and non-place neurons in the hippocampal network.   {\em Neuron} {\bf 96,} 1178--1191 (2017)
%
\bibitem{bialek+al_12}
W Bialek, A Cavagna, I Giardina, T Mora, E Silvestri, M Viale, and A Walczak,  Statistical mechanics for natural flocks of birds.   {\em Proc Natl Acad Sci (USA)} {\bf 109,} 4786--4791 (2012).
%
\bibitem{bialek+al_14}
W Bialek, A Cavagna, I Giardina, T Mora, O Pohl, E Silvestri, M Viale, and  AM Walczak, Social interactions dominate speed control in poising natural flocks near criticality.   {\em Proc Natl Acad Sci (USA)} {\bf 111,} 7212--7217 (2014).
%
\bibitem{mora+al_15}
 T Mora, S Deny, and O Marre, Dynamical criticality in the collective activity of a population of retinal neurons. {\em Phys Rev Lett} {\bf 114,} 078105 (2015).
%
\bibitem{cavagna+al_14a}
A Cavagna, I Giardina, F Ginelli, T Mora, D Poviani, R Tavarone, and AM Walczak, Dynamical maximum entropy approach to flocking.  {\em Phys Rev E}  {\bf 89,} 042707 (2014).
%
\bibitem{mora+al_16}
 T Mora, AM Walczak, L Del Castello, F Ginelli, S Melillo, L Parisi, M Viale, A Cavagna, and I Giardina, Local equilibrium in bird flocks. {\em Nature Physics} {\bf 12,} 1153 (2016).
%
\bibitem{ferrretti+al_20}
F Ferretti, V Chardes, T Mora, AM Walczak, and I Giardina, Building general Langevin models from discrete data sets.  {\em Phys Rev X} {\bf 10,} 031018 (2020).
%
\bibitem{tkacik+al_15}
G Tka\v{c}ik, T Mora, O Marre, D Amodei, SE Palmer, MJ Berry II, and W Bialek, Thermodynamics  and signatures of criticality in  a network of neurons.  {\em Proc Natl Acad Sci (USA)} {\bf 112,}  11508--11513 (2015).
%
\bibitem{ruelle_68}
D Ruelle, Statistical mechanics of a one-dimensional lattice gas. {\em Comm Math Phys} {\bf 9,} 267--278 (1968).
%
\bibitem{dyson_69}
FJ Dyson, Existence of a phase-transition in a one-dimensional Ising ferromagnet.  {\em Comm Math Phys} {\bf 12,} 91--107 (1969).
%
\bibitem{yuval+anderson_70}
G Yuval and PW Anderson, Exact results for the Kondo problem: One-body theory and extension to finite temperature.  {\em Phys Rev B} {\bf 1,} 1522--1528 (1970).
%
\bibitem{yuval+al_70}
G Yuval, PW Anderson, DR Hamman,  and Exact results for the Kondo problem. II. Scaling theory, qualitatively correct solution, and some new results on one-dimensional classical statistical models.  {\em Phys Rev B} {\bf 1,} 4464--4473 (1970).
%
\bibitem{schrodinger_44}
E Schr\"odinger, {\em What is Life?} (Cambridge University Press, Cambridge UK, 1944).
%
\bibitem{hopfield_74}
JJ Hopfield,  Kinetic proofreading: A new mechanism for reducing errors in biosynthetic processes requiring high specificity. {\em Proc Natl Acad Sci (USA)} {\bf 71,} 4135--4139 (1974).
%
\bibitem{ninio_75}
J Ninio, Kinetic amplification of enzyme discrimination. {\em Biochimie} {\bf  57,} 587--595 (1975).
%
\bibitem{sturtevant_13}
AH Sturtevant,   The linear arrangement of six sex--linked factors in {\em Drosophila}, as shown by their mode of association. {\em J Exp Zool} {\bf 14,} 43--59 (1913).
%
\bibitem{hodgkin+huxley_52}
AL Hodgkin and AF Huxley, A quantitative description of membrane current and its application
to conduction and excitation in nerve.  {\em J Physiol (Lond)} {\bf 117,} 500--544 (1952).
%
\bibitem{fatt+katz_52}
P Fatt and B Katz, Spontaneous subthreshold activity at motor nerve endings.  {\em J Physiol (Lond)} {\bf 117,} 109--128 (1952).
%
\bibitem{zweig_80}
G Zweig, Origins of the quark model.   In {\em Proceedings of the Fourth International Conference on Baryon Resonances,}  N Isgur, ed, pp 439--479 (University of Toronto, 1980). 
%
\bibitem{zweig_10}
G Zweig, Memories of Murray and the quark model.  {\em  Int J Mod Phys A} {\bf 25,} 3863--3877 (2010).
%
\bibitem{zweig_14}
G Zweig, Concrete quarks: The beginning of the end. {\em EPJ Web of Conferences} {\bf 71,} 00146 (2014).
%
\bibitem{bialek+al_07}
W Bialek, RR de Ruyter van Steveninck, and N Tishby, Efficient representation as a design principle for neural coding and computation.
 arXiv:0712.4381 [q--bio.NC] (2007).
 %
\bibitem{bouman_61}
MA Bouman, History and present status of quantum theory in vision.  In {\em Sensory Communication}, W Rosenblith, ed, pp. 377--401 (MIT Press, Cambridge MA, 1961).
%
\bibitem{rayleigh_07}
Lord Rayleigh, XII. On our perception of sound direction.  {\em Phil Mag Series 6} {\bf 13,} 214--232 (1907).
%
\bibitem{frisch_74}
K von Frisch, Decoding the language of the bee. {\em Science} {\bf 185,} 663--668 (1974).
%
\bibitem{gould_82}
JL Gould, {\em Ethology:  The Mechanisms and Evolution of Behavior} (WW Norton, New York, 1982).
%
\bibitem{hecht+al_42}
S Hecht, S Shlaer, and MH Pirenne, Energy, quanta and vision.  {\em J Gen Physiol}
{\bf 25,} 819--840 (1942).
%
\bibitem{bialek_12}
W Bialek, {\em Biophysics: Searching for Principles}   (Princeton University Press, Princeton NJ, 2012).
%
\bibitem{lawson+uhlenbeck_50}
JL Lawson and GE Uhlenbeck, {\em Threshold Signals.}  MIT Radiation Laboratory Series 24.  (McGraw--Hill, New York, 1950).
%
\bibitem{green+swets_66}
DM Green and JA Swets, {\em Signal Detection Theory and Psychophysics} (Wiley,
New York, 1966).
%
\bibitem{bouman_61}
MA Bouman, History and present status of quantum theory in vision.  In {\em Sensory Communication}, W Rosenblith, ed, pp. 377--401 (MIT Press, Cambridge MA, 1961).
%
\bibitem{rayleigh_07}
Lord Rayleigh, XII. On our perception of sound direction.  {\em Phil Mag Series 6} {\bf 13,} 214--232 (1907).
%
\bibitem{frisch_74}
K von Frisch, Decoding the language of the bee. {\em Science} {\bf 185,} 663--668 (1974).
%
\bibitem{gould_82}
JL Gould, {\em Ethology:  The Mechanisms and Evolution of Behavior} (WW Norton, New York, 1982).
%
\bibitem{hecht+al_42}
S Hecht, S Shlaer, and MH Pirenne, Energy, quanta and vision.  {\em J Gen Physiol}
{\bf 25,} 819--840 (1942).
%
\bibitem{bialek_12}
W Bialek, {\em Biophysics: Searching for Principles}   (Princeton University Press, Princeton NJ, 2012).
%
\bibitem{lawson+uhlenbeck_50}
JL Lawson and GE Uhlenbeck, {\em Threshold Signals.}  MIT Radiation Laboratory Series 24.  (McGraw--Hill, New York, 1950).
%
\bibitem{brunton+al_13}
BW Brunton, MW Botvinick, and CD Brody, Rats and humans can optimally accumulate evidence for decision-making. {\em Science} {\bf 340,} 95--98 (2013).
%
\end{thebibliography}
\end{document}